# Engineering Enhanced Thermal Transport in Layered Nanomaterials

Abhinav Malhotra[1], Kartik Kothari[2], and Martin Maldovan[1,2,*]

[1]*School of Chemical & Biomolecular Engineering, Georgia Institute of Technology, Atlanta GA 30332 USA*
[2]*School of Physics, Georgia Institute of Technology, Atlanta GA 30332 USA*

ABSTRACT:

A comprehensive rational thermal material design paradigm requires the ability to reduce and enhance the thermal conductivities of nanomaterials. In contrast to the existing ability to reduce the thermal conductivity, methods that allow to enhance heat conduction are currently limited. Enhancing the nanoscale thermal conductivity could bring radical improvements in the performance of electronics, optoelectronics, and photovoltaic systems. Here, we show that enhanced thermal conductivities can be achieved in semiconductor nanostructures by rationally engineering phonon spectral coupling between materials. By embedding a germanium film between silicon layers, we show that its thermal conductivity can be increased by more than 100% at room temperature in contrast to a free standing thin-film. The injection of phonons from the cladding silicon layers creates the observed enhancement in thermal conductivity. We study the key factors underlying the phonon injection mechanism and find that the surface roughness and layer thicknesses play a determining role. The findings presented in this letter will allow for the creation of nanomaterials with an increased thermal conductivity.

The ability to reduce heat conduction at nanoscales has been propelled by advances in the understanding of phonon transport processes in semiconductors [1-17]. In recent years, suppression of thermal conduction by orders of magnitude has been achieved through the diffuse scattering of phonons at nanostructure surfaces [1-8, 10, 11, 15]. Nanowires, superlattices, polycrystals, and nanocomposites with extremely low thermal conductivities have been introduced [1-8, 10, 11, 15, 18-25], which find applications as efficient thermoelectric materials. To date, nanostructuring has only been used to reduce the thermal conductivities for a wide range of temperatures. However, under a rational material design paradigm to create thermal materials, it is paramount to possess the ability to achieve both very low and very high thermal conductivities. Accessing the ability to enhance thermal conduction at nanoscales requires the development of novel approaches based on a deep understanding of phonon transport behavior. In this letter, we show that thermal conductivities at the nanoscale can also be enhanced. By carefully designing a tri-layer superlattice nanostructure and using two silicon layers as the outer cladding, the thermal conductivity of a germanium film (Figure 1a) can be increased by embedding the thin film between the silicon layers (Figure 1b). The enhancement is a consequence of the coupling of the phonon spectrum of germanium and silicon materials through the interfaces. Our focus is to elucidate the fundamental mechanism underlying this novel phenomenon and provide a guided design space for the experiments to enable the development of enhanced thermal conductivity in nanostructured materials. The latter has motivated the choice of materials, layer thicknesses, and use of experimentally quantifiable surface properties in this letter. Our results lay the foundation for the development of a fundamentally new avenue of thermal energy control at the nanoscale using phonon spectral coupling which may find significant applications in thermal design for highly-efficient heat dissipation in microelectronic, optoelectronic, and photovoltaic devices.

Thermal energy in semiconductors is carried by a broadband of phonons with different wavelengths and mean free paths[24, 26, 27]. In a tri-layer structure, if the characteristic sizes of two semiconductor materials are larger than the dominant mean free paths of heat-carrying phonons, the thermal conductivities of the materials will not be significantly perturbed by the presence of the other material, i.e. the materials will not "sense" each other when in contact. On the other hand, at the nanoscale, when the characteristic sizes are smaller than the phonon mean free paths, different layers of semiconductor materials will start to interact with each other through the interface. To model these interactions and the associated thermal conductivity of the three-layer system made of two different materials (Figure 1b) we use the well-established Boltzmann transport equation[28] $\partial f_{\mathbf{k},\alpha}/\partial t + \mathbf{v}_{\mathbf{k},\alpha} \cdot \nabla f_{\mathbf{k},\alpha} = -g_{\mathbf{k},\alpha}/\tau_{\mathbf{k},\alpha}$ where $f_{\mathbf{k},\alpha} = f_{\mathbf{k},\alpha}^0 + g_{\mathbf{k},\alpha}$ and $\alpha$ is the phonon polarization (i.e. longitudinal and transverse), $f^0=1/[\exp(\hbar\omega_{\mathbf{k}}/k_bT)-1]$ is the equilibrium distribution function, and $g_{\mathbf{k},\alpha}$ is the correction to the distribution function due to phonon scattering and the external thermal gradient $\nabla_x T$. All phonon distributions are functions of phonon frequency and direction of propagation, as indicated by the subscript wave-vector **k.** Additionally, the internal relaxation times $\tau_{\mathbf{k},\alpha}$[19] and propagation velocities $\mathbf{v}_{\mathbf{k},\alpha}$ are frequency dependent owing to the broadband nature of thermal phonons in semiconductors [24, 26, 27].

The mutual interaction between phonons of two materials is based on the ability of phonons to be transmitted at the interface. In general, upon interacting with a surface, a fraction of phonons undergo specular reflection and transmission, while the rest are randomized along all angular directions. We leverage this mutual exchange of phonons via transmission, to increase the thermal conductivity of a germanium thin film embedded between silicon layers with respect to a free-standing germanium thin film with the same physical properties. The cladding of silicon layers around the germanium thin-film act as a pump of phonons that can inject phonons into the germanium layer, allowing for an enhancement of its thermal conductivity due to their large transport properties (i.e. relaxation times and velocities). The efficiency of the phonon injection mechanism is controlled by the resistance to phonon transmission across the layers and the thicknesses of the layers. The resistance depends on the acoustic mismatch between the material pair, the diffuse scattering of phonons at interfaces, and the degree of non-overlap between phonon dispersion relations. The acoustic mismatch and the non-overlap of dispersion relations are intrinsic properties of the cladding and embedded material (i.e. silicon and germanium, respectively) and are fixed for a material pair. The diffuse scattering of phonons at interfaces, however, is determined by the properties of the interfaces (e.g. roughness and correlation lengths) between the cladding and internal layer. At the outer surfaces, the distribution function $g_{\mathbf{k}}$ for a phonon leaving the surface is equal to the distribution function for the phonon striking the surface multiplied by the probability of specular surface scattering $p(\mathbf{k},\theta,\eta,L)$ where $\theta$ is the incident angle, $\eta$ is the surface roughness and $L$ is the correlation length[29]. On the other hand, at the silicon/germanium interface, the distribution function contains contributions from reflection and transmission events with probabilities $P(\mathbf{k},\theta,\eta,L)$ and $Q(\mathbf{k},\theta,\eta,L)$, respectively[30]. Note that these probabilities are a function of surface and incident phonon properties i.e. surface roughness, correlation length, incident phonon momentum and angle. We use a generalized analysis of surface scattering model stemming from the Beckmann-Kirchhoff theory[31] that includes forward scattering (i.e. reflection and transmission) of phonons at a rough interface between two solid materials. The reflection $P_{ij}$ and transmission $Q_{ij}$ coefficients from a rough interface, where $i$ and $j$ denote the two solid media across the surface, can be written as $P_{ij} = R_{ij}^2 \exp(-4\eta^2 k_i^2 \cos^2\theta_i)$ and $Q_{ij} = (1-R_{ij}^2)\exp[-\eta^2(k_i\cos\theta_i - k_j\cos\theta_j)^2]$, where $\theta_i$ and $\theta_j$ are the incident and transmitted angles respectively, and $R_{ij}$ is the specular coefficient as given by Fresnel equations[30]. These expressions are extended for surface shadowing to account for the undulating nature of the surface[29]. Using these probabilities we perform an energy balance at the inner interfaces (Equation 1) to obtain the functions $g_\mathbf{k}$ for each material layer.

$$\begin{cases} g_{\mathbf{k},1}^- = P_{12}(\mathbf{k},\theta,\eta,L)\, g_{\mathbf{k},1}^+ + Q_{12}(\mathbf{k},\theta,\eta,L)\, g_{\mathbf{k},2}^- \\ g_{\mathbf{k},2}^+ = P_{21}(\mathbf{k},\theta,\eta,L)\, g_{\mathbf{k},2}^- + Q_{12}(\mathbf{k},\theta,\eta,L)\, g_{\mathbf{k},1}^+ \end{cases} \quad (at\ z=t_{Si})$$

$$\begin{cases} g_{\mathbf{k},2}^- = P_{23}(\mathbf{k},\theta,\eta,L)\, g_{\mathbf{k},2}^+ + Q_{23}(\mathbf{k},\theta,\eta,L)\, g_{\mathbf{k},3}^- \\ g_{\mathbf{k},3}^+ = P_{32}(\mathbf{k},\theta,\eta,L)\, g_{\mathbf{k},3}^- + Q_{23}(\mathbf{k},\theta,\eta,L)\, g_{\mathbf{k},2}^+ \end{cases} \quad (at\ z=t_{Si}+t_{Ge}) \quad (1)$$

The superscript $+$ and $-$ denote the direction of propagation determined by positive and negative $v_{z,\mathbf{k}}$. The subscripts $i=1,2,3$ denote the layer number in the tri-layer structure with the embedded germanium corresponding to $i=2$. The thermal flux and the associated thermal conductivity of the germanium film $\kappa_{Ge}$ is calculated by integrating the distribution function over all phonon wavevectors $\mathbf{k}$ for all polarizations (i.e. two transverse and one longitudinal):

$$\mathbf{J}_{Ge} = -\kappa_{Ge}\nabla T = \frac{1}{t_{Ge}}\int_0^{t_{Ge}}[\frac{1}{(2\pi)^3}\sum_\alpha \int \hbar\omega_{\mathbf{k},\alpha} f_{\mathbf{k},\alpha} \mathbf{v}_{\mathbf{k},\alpha} d^3k]dz \quad (2)$$

The calculated thermal conductivities of a germanium thin film in the tri-layer structure are shown in Figures 2a and 2b as a function of the thickness of the silicon cladding layer, where a significant enhancement in thermal conductivity of the germanium film over the free-standing counterpart is observed. We find that it is possible to nearly double the thermal conductivity (>90% enhancement) of a 10nm germanium thin-film by using two 1μm-thick silicon samples as cladding. The specularities of the cladding-air interface are considered to be $p=0$ and $p=1$ to cover all possibilities in terms of the quality of the silicon-air interfaces [28]. The germanium-silicon interface has roughness $\eta=0.1$nm. A germanium thin-film with given physical properties (surface roughness, correlation length, and thickness) without any cladding is used as the baseline measure for comparison. From the graphs in Figures 2a and 2b, it is clear that the larger the cladding thickness, the larger the increase in the thermal conductivity of the germanium thin film. This is because, for larger thicknesses of the silicon cladding layers, phonon scattering at the interfaces is reduced and silicon phonon mean free paths are not shortened previously to their injection in germanium. Note that the enhancement of conductivity via injection of phonons is not unbounded with increasing cladding thickness, rather it begins to saturate as shown in Figure 2a. Clearly, increasing cladding thickness beyond the saturation limit is of no additional advantage from the perspective of enhancement, as the thermal conductivity contribution of phonons that can interact with the embedded layer has already been maximized. The specularity of the cladding-air interface is another factor that can influence the injection of phonons into the embedded layer. At the same cladding thickness, the observed enhancement of thermal conductivity in the embedded layer is higher for larger cladding-air surface specularities. This behavior is consistent since a more diffuse cladding shortens the phonon mean-free-paths, and therefore a larger thickness is required to improve the injection efficiency. Note that at sufficiently large cladding thicknesses, the impact of the quality of the silicon-air interface on germanium thermal conductivity enhancement begins to diminish and in the case of a bulk silicon cladding it would be expected that the enhancement is the same irrespective of the roughness of the cladding-air surface. Note that coherent modifications to the phonon dispersion relations can be neglected at room temperature owing to the choice of embedded layer thicknesses[32].

Figures 3a and 3b show the impact of silicon-germanium interface roughness on the increased thermal conductivity of the germanium film. Significant enhancements (~100%) in thermal conductivities of the 10nm embedded layer with roughness values in range of $\eta$=0.1-0.4nm can be achieved by using 1μm silicon layers as cladding. Additionally, with increasing thickness of the embedded germanium layer, the maximum enhancement is reduced as the cladding is only able to inject phonons into a part of the thicker embedded germanium film. That is, the proportion of region in the germanium layer that is able to augment its local conductivity with the injected phonons from silicon decreases, leading to a smaller enhancement in thermal conductivity. Another interesting observation is that the thermal conductivity of a germanium thin-film can be enhanced beyond the bulk thermal conductivity of germanium at room temperature (60W/m-K) using the described tri-layer architecture. However, it is important to note that the thermal conductivity enhancement in the embedded-layer occurs at the cost of reducing the thermal conductivity of the silicon cladding layer (as compared to a baseline silicon film of similar physical properties). In simple words, the cladding layers allow for a local enhancement of the thermal conductivity in the embedded layer at the cost of reducing their own thermal conductivity.

To further explain the origin of the thermal conductivity enhancement, we analyze the transport behavior of a well-established system of a "nanoparticle-in-alloy" bulk semiconductor[4, 7] and compare it with our nanostructured semiconductor. In a bulk structure, the thermal conductivity at a point $O$ in real space is determined on an average by phonons coming from a spherical surface centered at that point with radius equaling the bulk mean free path in that semiconductor. Any changes to the semiconductor structure (such as addition of alloy atoms and nanoparticles) outside this "influence-sphere" can be considered to minimally affect the thermal conductivity at the point $O$. On the other hand, the inclusion of alloy atoms and nanoparticles within this influence-sphere will reduce the thermal conductivity by shortening the bulk phonon mean free paths, thereby reducing the effective radius of the influence-sphere. Analogously to the bulk case, in the case of a free standing thin-film (Figure 4a), the thermal conductivity at a point $O$ inside the film is determined on an average by phonons arriving from the surface of the influence sphere (see black circular line) given by the thin-film mean free path. We analyze the phonon trajectories and the impact of scattering on the effective phonon mean free paths for the germanium thin film and the tri-layer superlattice structure in Figure 4. The addition of the silicon cladding layers to the germanium thin film (Figure 4b) ensures that in addition to the phonons within the germanium layer, transmitted phonons from silicon that can couple across the interface and reach point *O* are able to enhance the thermal conductivity by increasing the effective mean free paths (red circle). This is because phonons arriving from silicon through transmission had their last collision on an average at a larger distance than those arriving from germanium. Consequently, we show that nanostructuring does not necessarily lead to a reduction in the germanium thermal conductivity, alternatively it can generate an enhancement of heat conduction.

In summary, we have shown that with engineered nanostructuring, the thermal conductivity of nanomaterials can be enhanced. We showed that the use of a tri-layer silicon-germanium-silicon structure allows for coupling phonon spectra across interfaces and can increase the thermal conductivity of the embedded germanium layer. A significant enhancement in thermal conductivity (~2×) of an embedded germanium layer above its free-standing value can be engineered under specific surface conditions and layer sizes. Under certain conditions, it may also be possible to increase the thermal conductivity of the germanium layer beyond the bulk value. A smoother interface between the two material layers and a larger thickness of cladding silicon layers was found to promote the thermal conductivity enhancement. In general, the effect of enhancement will be controlled by the choice of material pairs, the sizes of specific layers and the condition of interfaces. Thermal conductivity enhancement via coupled phonon transport as presented in this letter has the potential to revolutionize the rational thermal material design paradigm by generating the ability to increase the thermal conductivity in nanostructures. The leap beyond the limited use of nanostructuring to only reduce thermal transport via the use of phonon spectral coupling significantly advances the fundamental understanding and control of thermal transport in nanomaterials.

**Figure Captions**

**Figure 1.** Tri-layer superlattice nanostructure. (a) Schematics for a free-standing germanium thin film (yellow). (b) By embedding the germanium thin film between silicon layers (orange) the thermal conductivity of the germanium film can be increased. The temperature gradient is applied along the $x$ direction.

**Figure 2.** Increased thermal conductivity. (a) Colored areas show the enhancement in the thermal conductivity of a germanium thin film as a function of increasing thickness of the silicon cladding layer. The top and bottom lines correspond to silicon-air surface specularities $p=1$ and $p=0$, respectively, covering all possibilities in terms of surface roughness. The germanium-silicon interface has roughness $\eta=0.1$nm and correlation length $L=20\eta$. (b) Enhanced thermal conductivity of an embedded germanium film (solid lines) with respect to a free-standing germanium thin film (dotted lines) having the same physical properties. The germanium-silicon interface has $\eta=0.1$nm, $L=20\eta$ while the silicon-air surface specularity is equal to one. The temperature gradient is in the plane of the films.

**Figure 3.** Enhanced heat conduction. Thermal conductivity increase for the embedded germanium film as a function of the silicon-germanium interface roughness. The germanium thin film thickness is (a) $t_{Ge}=10$nm and (b) $t_{Ge}=100$nm. Color lines correspond to different thicknesses of the silicon layers, ranging from no silicon layer to bulk silicon. The enhancement in the thermal conductivity of the embedded germanium film, with respect to the free-standing germanium thin film with same properties, is observed for various surface roughnesses and silicon layer thicknesses.

**Figure 4.** Phonon Injection. (a) Schematic for phonon contributions to the thermal conductivity of a germanium thin film under a temperature gradient along the $x$ direction. The thermal flux at point $O$ is carried by phonons whose last collision was, on the average, at a distance of mean free path $\ell_{TF}$ away from $O$, as represented by the black circular line. For simplicity we neglect the angular dependence of $\ell_{TF}$. (b) When the germanium film is in contact with the silicon film, phonons from silicon can be injected into germanium (blue arrow). These phonons had their last collision at a larger distance than those arriving from germanium. As a result, the effective mean free path $\ell_{EFF}$ is larger (red circular line) and the thermal conductivity at point $O$ is enhanced.

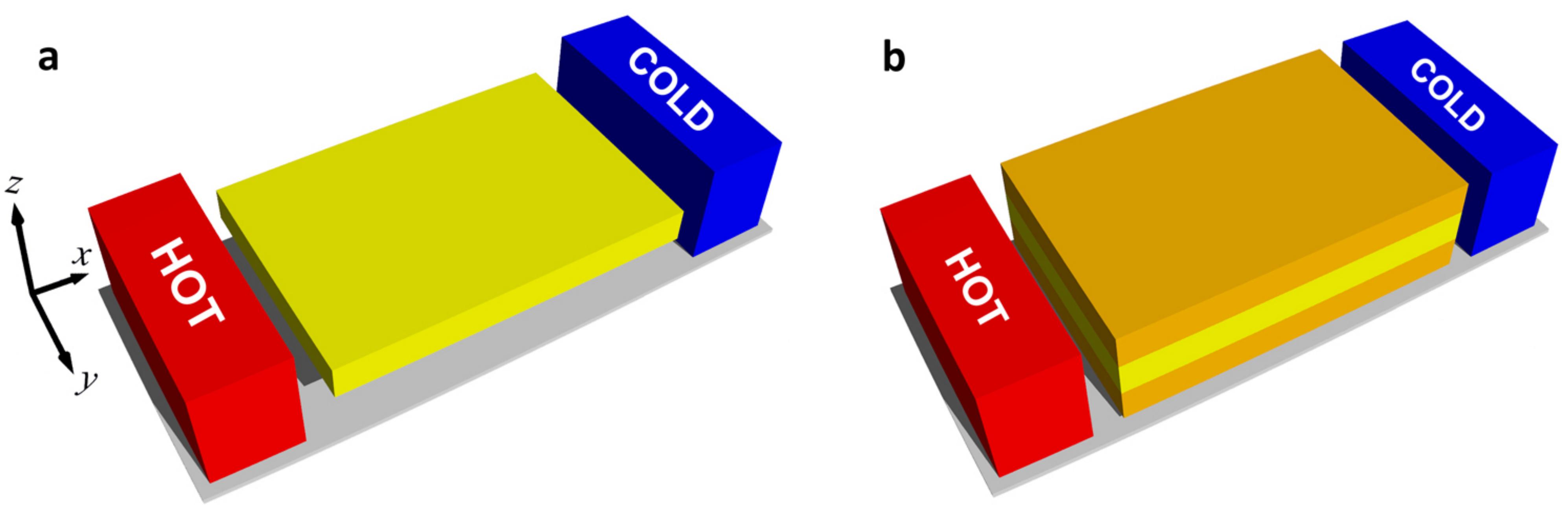

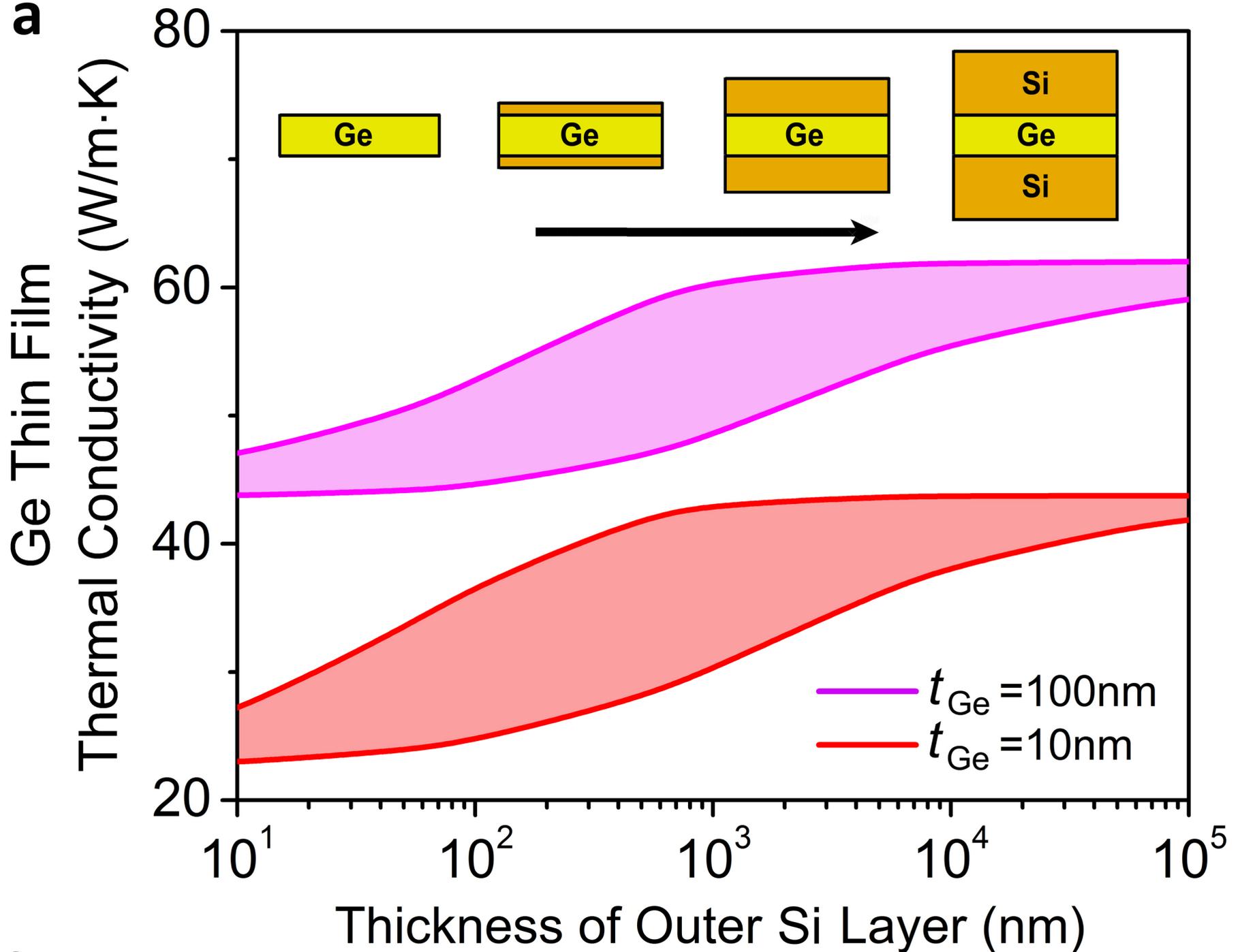
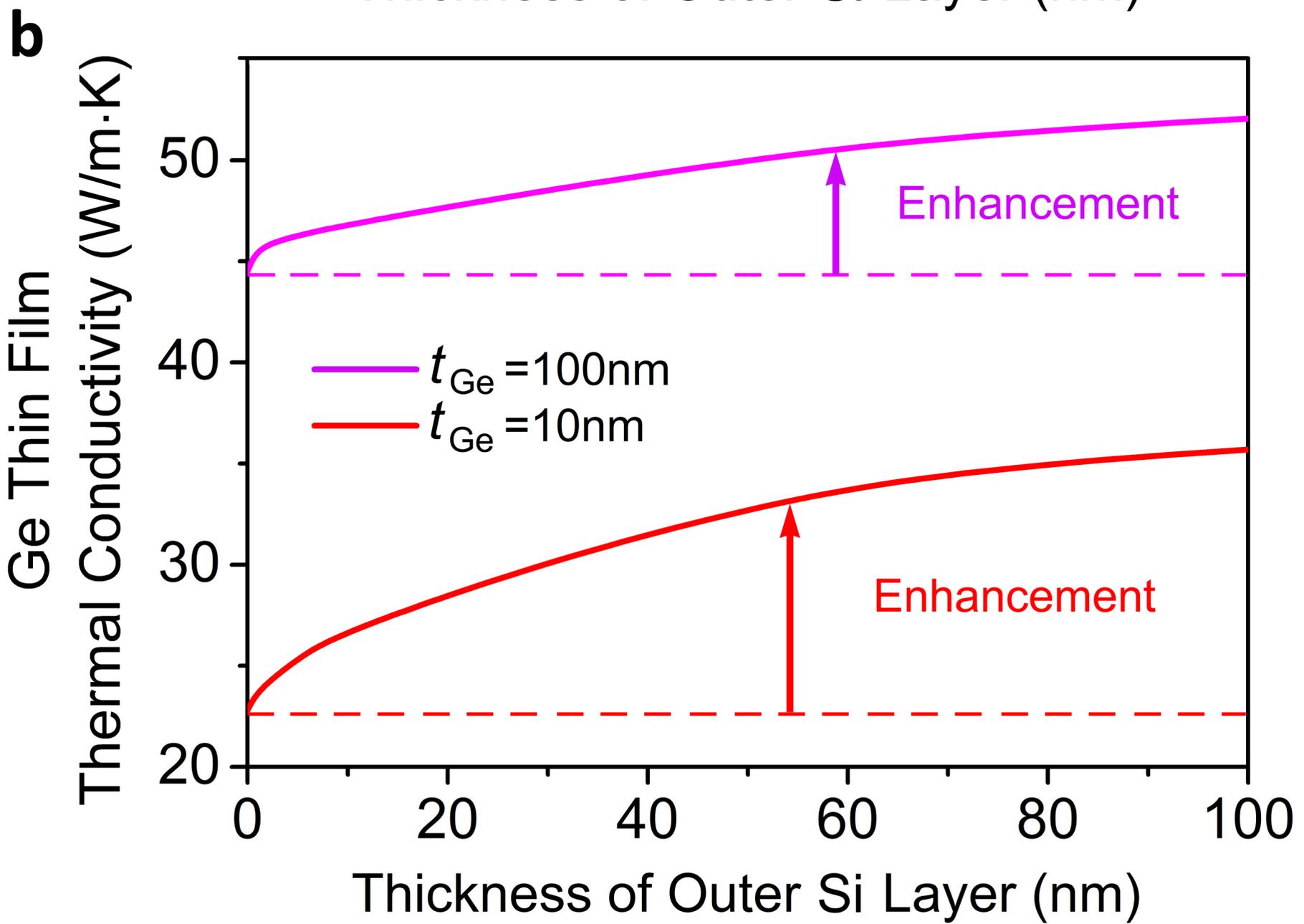

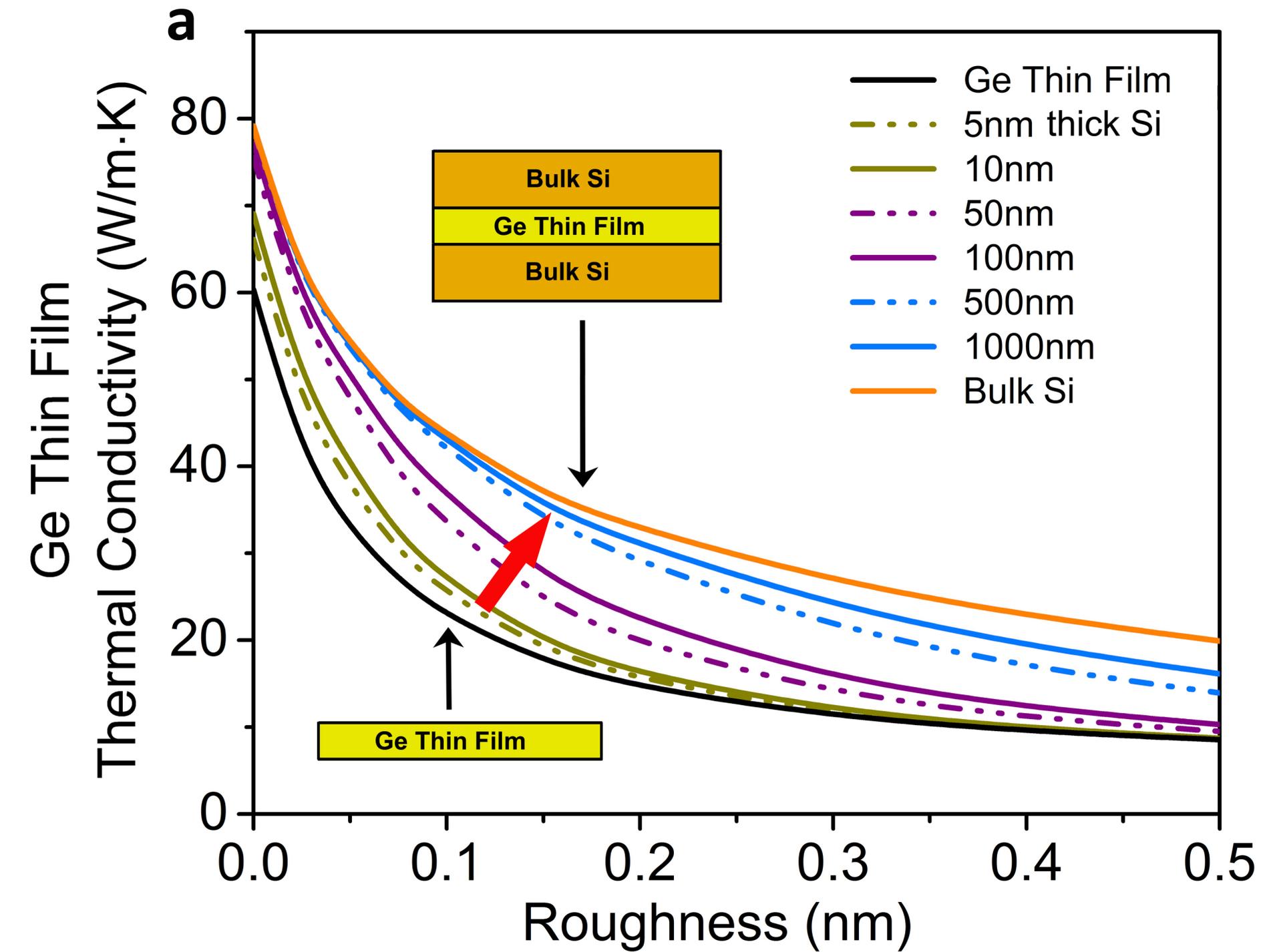 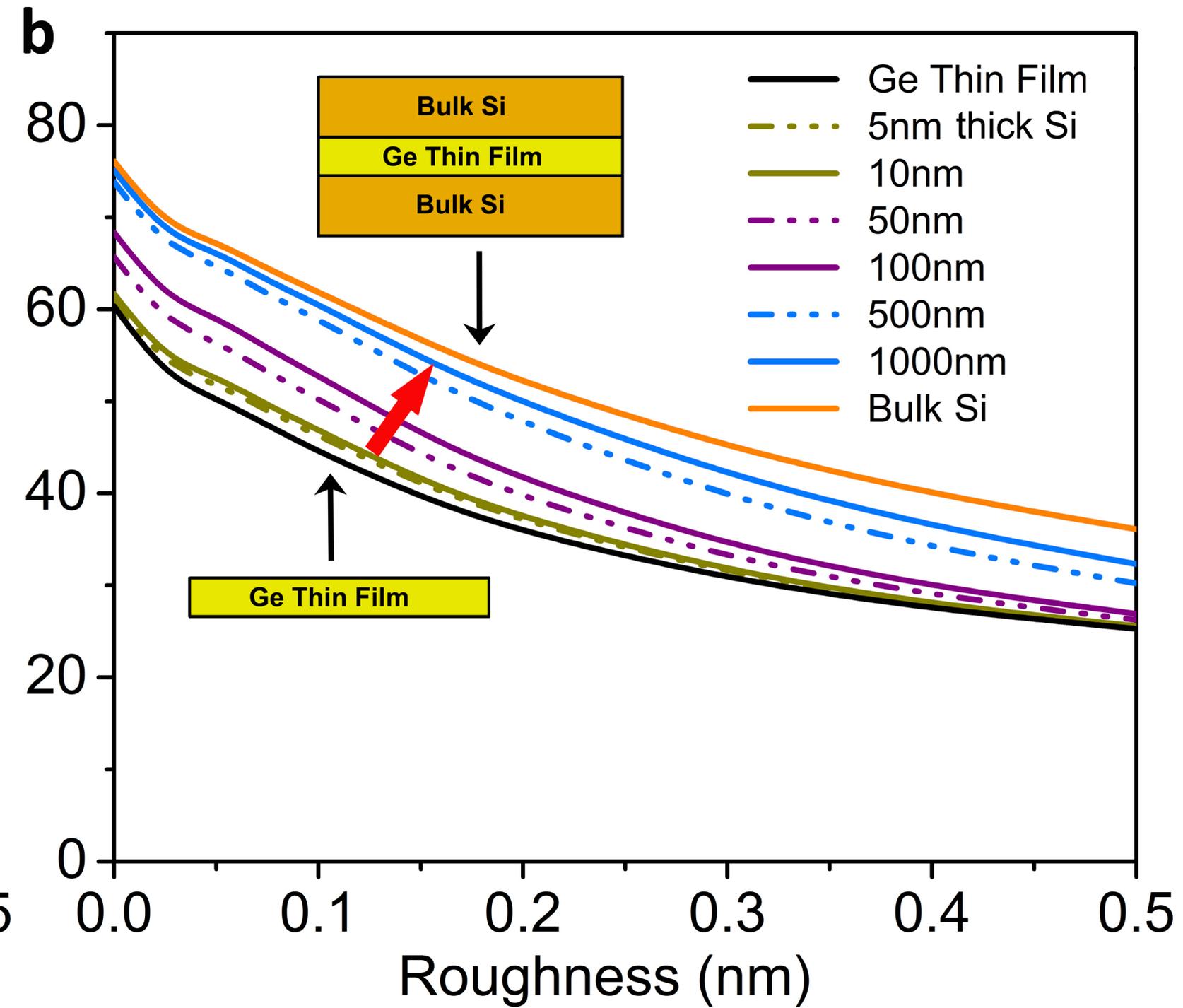

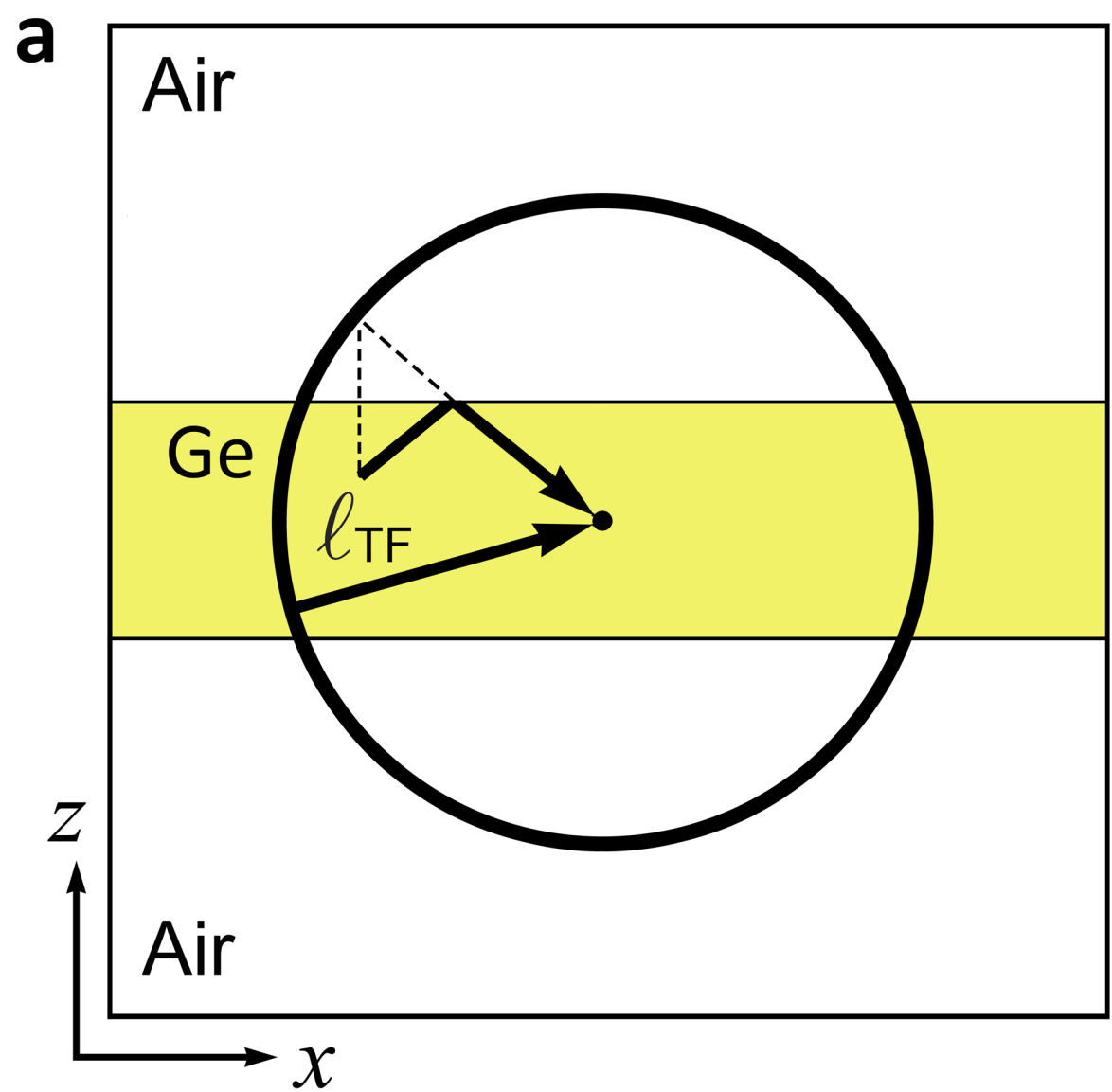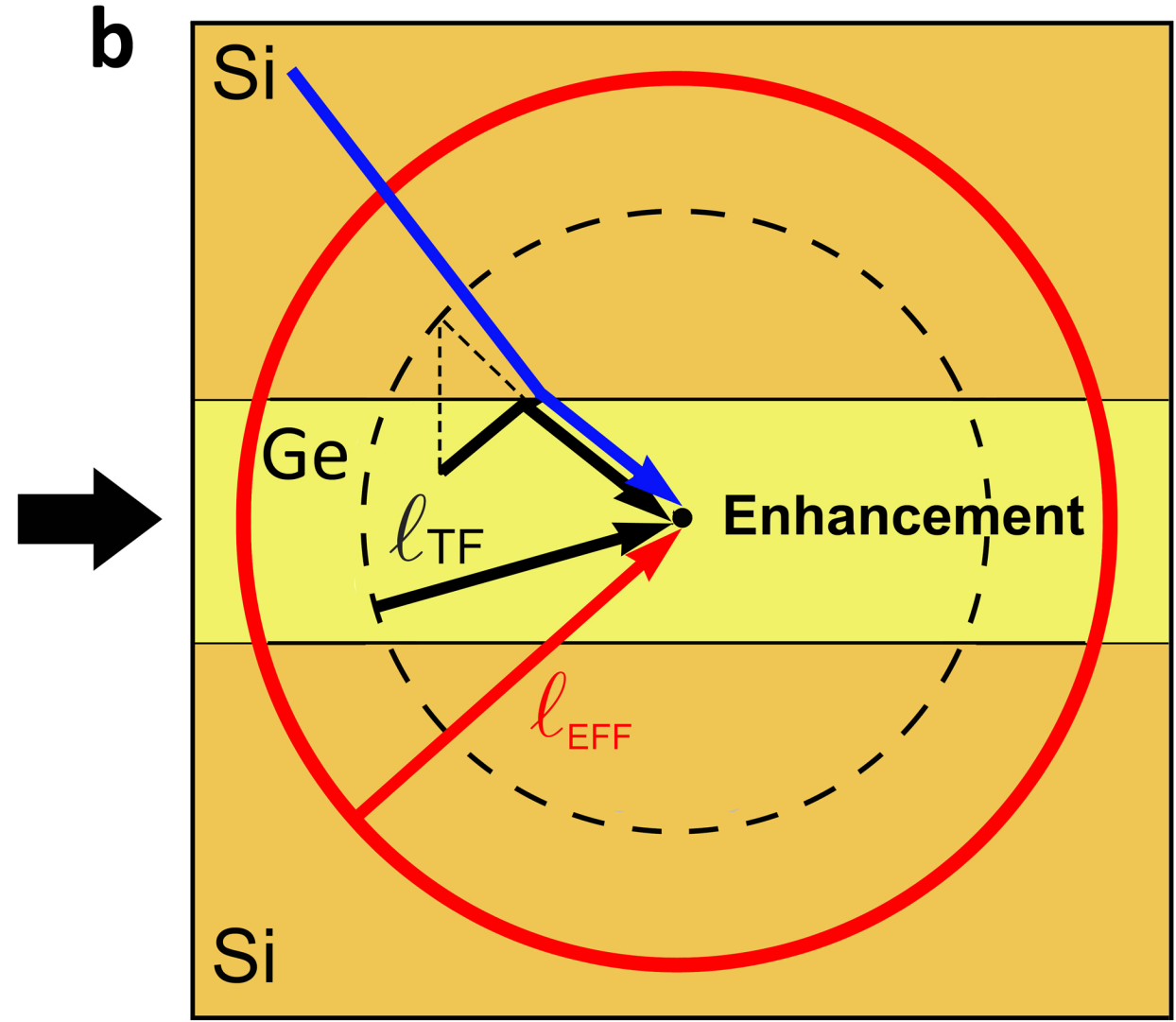